\documentclass[11pt]{article}
\usepackage{latexsym}
\usepackage{amsmath} 
\usepackage{amscd}     \usepackage{amsxtra}
\usepackage{upref}     \usepackage{amsthm}
\usepackage{amssymb}   
\usepackage{amsfonts}
\begin{document}

\title{{\bf On the Structure of the Nonlinear Vacuum \\Solutions in Extended
Electrodynamics}}

\author{{\bf Stoil Donev}\footnote{e-mail:
 sdonev@inrne.bas.bg} \\ Institute for Nuclear Research and Nuclear
Energy,\\ Bulg.Acad.Sci., 1784 Sofia, blvd.Tzarigradsko chaussee 72,
Bulgaria\\
{\bf Maria Tashkova}\\Institute of Organic Chemistry with Center of
Phytochemistry,\\ Bulg. Acad. Sci., 1113 Sofia, Acad.
G.Bonchev Str.  9, Bulgaria}

\date{}
\maketitle

\begin{abstract}
In this paper, in the frame of Extended Electrodynamics (EED), we study some
of the consequences that can be obtained from the introduced and used by
Maxwell equations  complex structure $\mathcal{J}$ in the space of 2-forms on
$\mathbb{R}^4$, and also used in EED.  First we give the vacuum EED equations
with some comments. Then we recall some facts about the invariance group $G$
(with Lie algebra $\mathcal{G}$) of the standard complex structure $J$ in
$\mathbb{R}^2$. After defining and briefly studying a representation of $G$
in the space of 2-forms on $\mathbb{R}^4$ and the joint action of $G$ in the
space of the $\mathcal{G}$-valued 2-forms on $\mathbb{R}^4$ we consider its
connection with the vacuum solutions of EED.  Finally, we consider the case
with point dependent group parameters and show that the set of the nonlinear
vacuum EED-solutions is a disjoint union of orbits of the $G$-action, noting
some similarities with the quantum mechanical eigen picture and with the QFT
creation and annihilation operators.
\end{abstract}


\section{Introduction}
Extended Electrodynamics (EED) [1] was brought to life in searching for
appropriate ways to describe {\it spatially-finite and time-stable}
electromagnetic field configurations propagating along a given direction in
the 3-space. In view of the adequate enough to reality energy-momentum
quantities and relations that follow from Maxwell theory, a basic requirement
to the modification of Maxwell equations was to keep the same local
energy-momentum quantities (i.e. the energy-momentum tensor) and relations.
The 4-dimensional final form of the vacuum EED-equations was assumed to be
\begin{equation}
\mathbf{d}F\wedge *F=0,\ \ \mathbf{d}*F\wedge F=0,\ \
F\wedge *\mathbf{d}*F+*F\wedge *\mathbf{d}F=0,
\end{equation}
or in component form
\[
F^{\mu\nu}(\mathbf{d}F)_{\mu\nu\sigma}=0,\ \
(*F)^{\mu\nu}(\mathbf{d} *F)_{\mu\nu\sigma}=0, \ \
(*F)^{\mu\nu}(\mathbf{d}F)_{\mu\nu\sigma}+
F^{\mu\nu}(\mathbf{d} *F)_{\mu\nu\sigma}=0,\ \ \ \mu<\nu,
\]
where $*$ is the Hodge operator defined by the MInkowski metric $\eta$ with
signature $(-,-,-,+)$ on $\mathbb{R}^4$, and $\mathbf{d}$ is the exterior
derivative.

We recall that Maxwell equations $\mathbf{d} F=0,\ \ \mathbf{d}*F=0$ describe
differentially the {\it balance} between the flows of the electric
$\mathbf{E}$ and magnetic $\mathbf{B}$ vectors through finite closed and
not-closed 2-surfaces. This approach to description of the propagation of
electromagnetic (EM) fields has, probably, been suggested by analogy with
hydrodynamics. The substantial difference, however, is in, that in
hydrodynamics the propagating vector quantities have a direct physical sense
of {\it mass-energy} and {\it momentum} densities, which are considered in
physics as {\it universal conserved quantities}, and this property allows to
write down corresponding differential (local) energy-momentum balance
relations.  As we noted, in electrodynamics the flows of $\mathbf{E}$ and
$\mathbf{B}$ have {\it not} a direct physical meaning of energy-momentum
flows.  This feature of the Maxwell approach to describe the field dynamics
resulted in establishing the linear character of the equations for
$F_{\mu\nu}$, and the D'Alembert linear evolution equation (i.e. $\square
U=0$) for each of the field components appeared as a necessary condition. As a
consequence, this lead to the impossibility to describe appropriately by
exact solutions to the vacuum Maxwell equations finite portions of
vacuum EM-radiation with soliton-like properties.

The 20th century physics proved, however, that the EM-radiation consists of
(almost) noninteracting photons, which we consider as finite time-stable
objects with soliton-like behavior, so, 3d soliton-like portions of
EM-radiation are possible to be created experimentally.  The soliton-like
evolution of these vacuum electromagnetic 3d soliton-like configurations
should have a consistent description even at non-quantum level.  In an
attempt to achieve this EED extended Maxwell equations to the nonlinear
equations (1).  In contrast to the Maxwell approach, equations (1) have a
{\it direct} physical sense of {\it local energy-momentum balance relations}.
Moreover, EED succeeded to keep the same basic energy-momentum expressions
and relations from Maxwell theory, and at the same time it extends seriously
the class of solutions.  So that, all vacuum solutions to Maxwell equations
are solutions to (1), but our equations (1) have more solutions satisfying
the relations $\mathbf{d}F\neq 0,\ \ \mathbf{d}*F\neq 0$.  These new
solutions to (1) will be called further {\it nonlinear}, and will be the
subject of study from a definite point of view.  We begin with recalling the
following result of EED specifying some of their properties.

\vskip 0.5cm
\noindent
{\bf For every nonlinear solution of (1) there exists a canonical
coordinate system \linebreak
$(x,y,z,\xi=ct)$ in the Minkowski space-time in which the
solution is fully represented by two functions: $\phi(x,y,\xi +\varepsilon
z),\  \varepsilon=\pm 1$, called {\it amplitude function}, and
$\varphi(x,y,z,\xi), |\varphi|\leq 1$, called {\it phase function}, in the
following way:
\begin{equation}
F=\varepsilon\phi \varphi dx\wedge dz +
\varepsilon\phi\sqrt{1- \varphi^2} dy\wedge dz+
\phi \varphi dx\wedge d\xi + \phi\sqrt{1- \varphi^2} dy\wedge d\xi.
\end{equation}
Moreover, every nonlinear solution satisfies also}:
\[
(\delta F)^2=(\delta *F)^2<0,\ \ (\delta F)_\sigma (\delta *F)^\sigma=0,\ \
F_{\mu\nu}F^{\mu\nu}=F_{\mu\nu}(*F)^{\mu\nu}=0.
\]
where $\delta=*\mathbf{d}*$ is the coderivative.
\vskip 0.5cm
\noindent
We recall that in this case the energy tensor
\[
Q_\mu^\nu=-\frac{1}{4\pi}\big[(F_{\mu\sigma}F^{\nu\sigma}+
(*F)_{\mu\sigma}(*F)^{\nu\sigma}\big]
\]
has just one isotropic eigen direction [2],
which is eigen direction for $F$ and $*F$
too, and in this coordinate system it is defined by the vector field
$\zeta=-\varepsilon\frac{\partial }{\partial z}+\frac{\partial}{\partial
\xi}$.

In order to come to the nonlinear equations (1) EED considers the fields
$(F,*F)$ as two components of a differential 2-form $\Omega=F\otimes
e_1+*F\otimes e_2$ on $\mathbb{R}^4$ with values in a 2-dimensional real
vector space $V$, which as a vector space can be identified with
$\mathbb{R}^2$. But a further study of the duality symmetry of the equations
suggested this vector space $V$ to be identified with the Lie algebra of the
corresponding Lie group of matrices of the kind
\begin{equation}
G=\left\{\begin{Vmatrix}a & -b \\ b & a\end{Vmatrix},    
\quad a,b\in \mathbb{R}\right\},
\end{equation}
which represents the dual symmetry of Maxwell vacuum equations. And further
we consider what consequences can be made from this suggestion.

\section{Some facts concerning the group $G$}
The group $G$ (3) is a commutative 2-dimensional Lie group with respect to
the usual product of matrices. In fact
\[
\begin{Vmatrix}a & -b \\ b & a\end{Vmatrix}.
\begin{Vmatrix}m & -n \\ n & m\end{Vmatrix}=
\begin{Vmatrix}m & -n \\ n & m\end{Vmatrix}.
\begin{Vmatrix}a & -b \\ b & a\end{Vmatrix}=
\begin{Vmatrix}am-bn & -(an+bm) \\ an+bm & am-bn\end{Vmatrix} .
\]
In addition, $G$ is a real 2-dimensional vector space with respect to the
usual addition of matrices. A natural basis of the vector space $G$ is given
by the two matrices
\[
I=\begin{Vmatrix}1 & 0\\0 & 1\end{Vmatrix},\
J=\begin{Vmatrix}0 & -1\\1 & 0\end{Vmatrix}.
\]
Since the Lie group $G$ is commutative, its Lie algebra $\mathcal{G}$ is
trivial, and as a vector space it coincides with the vector space $G$.
Hence, every element $\alpha$ of $G$ and $\mathcal{G}$ can be
represented as $\alpha=aI+bJ,\ a,b\in \mathbb{R}$. We have $J.J=-I$, so,
$J$ defines a complex structure in $\mathcal{G}$. If $\alpha\in G$
is given by $\alpha=aI+bJ$, then
\[
\alpha.J=-bI+aJ,\ \
\alpha^{-1}=\frac{1}{a^2+b^2}(aI-bJ),\ \
\alpha^{-1}.J=\frac{1}{a^2+b^2}(bI+aJ),\quad det(\alpha)=a^2+b^2
\]
The product of two matrices $\alpha=aI+bJ$ and $\beta=mI+nJ$
looks like $\alpha.\beta=(am-bn)I+(an+bm)J$. The
commutativity of $G$ means symmetry, in particular, every $\alpha\in G$ is a
symmetry of $J$:  $\alpha.J=J.\alpha$. One sees that $G$ is the well known
real matrix representation of the multiplicative group of the complex
numbers, the complex conjugation is given by $\bar\alpha=aI-bJ$, and further
we shall write $\alpha=aI+\varepsilon bJ$, $\varepsilon=\pm 1$.

\section{The action of $G$ in the space of 2-forms on $\mathbb{R}^4$}
We consider now the space $\mathbb{R}^4$ as a manifold, and denote by
$(x^1,x^2,x^3,x^4=x,y,z,\xi)$, where $\xi=ct$, the canonical coordinates of
$\mathbb{R}^4$, and by $\Lambda^2(\mathbb{R}^4)$ the space of 2-forms on
$\mathbb{R}^4$. As in [3] we introduce the following basis of
$\Lambda^2(\mathbb{R}^4)$:
\[
dx\wedge dy,\ \ dx\wedge dz,\ \ dy\wedge dz,\ \ dx\wedge d\xi,\ \
dy\wedge d\xi,\ \ dz\wedge d\xi.
\]
We recall from [3] that Maxwell equations
$\mathbf{d}F=0,\ \mathbf{d}\mathcal{J}F=0$ introduce just
the complex structure $\mathcal{J}$ in the space $\Lambda^2(\mathbb{R}^4)$,
and $\mathcal{J}$ acts on the above basis as follows:
\begin{align*}
\mathcal{J}(dx\wedge dy) &=-dz\wedge d\xi &
\mathcal{J}(dx\wedge dz) &=dy\wedge d\xi  &
\mathcal{J}(dy\wedge dz) &=-dx\wedge d\xi \\
\mathcal{J}(dx\wedge d\xi) &=dy\wedge dz &
\mathcal{J}(dy\wedge d\xi) &=-dx\wedge dz &
\mathcal{J}(dz\wedge d\xi) &=dx\wedge dy.
\end{align*}
Hence, in this basis the matrix of $\mathcal{J}$ is off-diagonal with
alternatively ordered $(-1,1,-1,1,-1,1)$.

Let now $\mathcal{I}$ is the identity map in $\Lambda^2(\mathbb{R}^4)$. We
define a representation $\rho$ of $G$ in $\Lambda^2(\mathbb{R}^4)$ as
follows:
\begin{equation}
\rho(\alpha)=
\rho(aI+\varepsilon bJ)=a\mathcal{I}+\varepsilon b\mathcal{J}.  
\end{equation}
So, the map $\rho$ is a linear map, it sends the elements of the linear space
$G$ to the space of linear maps $L_{\Lambda^2(\mathbb{R}^4)}$ of
$\Lambda^2(\mathbb{R}^4)$, so that every $\rho(\alpha)$ is a linear
isomorphism, in fact, its determinant $det||\rho(\alpha)||$ is equal to
$(a^2+b^2)^3$.  The identity $I$ of $G$ is sent to the identity $\mathcal{I}$
of $\Lambda^2(\mathbb{R}^4)$, and the complex structure $J$ of
the vector space $G$ is sent to the complex structure $\mathcal{J}$
of $\Lambda^2(\mathbb{R}^4)$. This map is surely a representation, because
$\rho(\alpha.\beta)=\rho(\alpha).\rho(\beta)$.  In fact,
\[
\rho(\alpha.\beta)=
\rho\big[(aI+\varepsilon bJ).(mI+\varepsilon nJ)\big]=
\]
\[
\rho\big[(am-bn)I+\varepsilon(an+bm)J\big]=
(am-bn)\mathcal{I}+\varepsilon (an+bm)\mathcal{J}.
\]
On the other hand
\[
\rho(\alpha).\rho(\beta)
=\rho(aI+\varepsilon bJ).\rho(mI+\varepsilon nJ)
\]
\[
=(a\mathcal{I}+\varepsilon b\mathcal{J}).
(m\mathcal{I}+\varepsilon n\mathcal{J})=
(am-bn)\mathcal{I}+\varepsilon (an+bm)\mathcal{J}.
\]

We consider now the space $\Lambda^2(\mathbb{R}^4,\mathcal{G})$
of $\mathcal{G}$-valued 2-forms on $\mathbb{R}^4$.
Every such 2-form $\Omega$ can be represented as
$\Omega=F_1\otimes I+F_2\otimes J$, where $F_1$ and $F_2$ are 2-forms. We
have the joint action of $G$ in $\Lambda^2(\mathbb{R}^4,\mathcal{G})$ as
follows:
\[
[\rho(\alpha)\otimes \alpha].\Omega=
\rho(\alpha).F_1\otimes \alpha.I+\rho(\alpha).F_2\otimes \alpha.J.
\]
We obtain
\[
[\rho(\alpha)\otimes \alpha].\Omega=
\]
\[
\big[(a^2\mathcal{I}+\varepsilon ab\mathcal{J})F_1-
(b^2\mathcal{J}+\varepsilon ab\mathcal{I})F_2\big]\otimes I+
\big[(b^2\mathcal{J}+\varepsilon ab\mathcal{I})F_1+
(a^2\mathcal{I}+\varepsilon ab\mathcal{J})F_2\big]\otimes J.
\]
In the special case $\Omega=F\otimes I+\mathcal{J}.F\otimes J$ it readily
follows that
\begin{equation}
[\rho(\alpha)\otimes \alpha].\Omega=(a^2+b^2)\Omega.             
\end{equation}
In this sense the forms $\Omega=F\otimes I+\mathcal{J}.F\otimes J$ may be
called {\it conformally equivariant} with respect to the action of the
group $G$, and {\it equivariant} with respect to the subgroup $SO(2)\subset
G$, i.e.  when $a^2+b^2=1$.

\vskip 0.3cm
\noindent
{\bf Remark}. The two partial actions will be denoted further by
$\rho(\alpha).\Omega$ and $\alpha.\Omega$. So, the above
conformal equivariance property may be written as
$$
\rho(\alpha).\Omega=\alpha^{-1}.(a^2+b^2)\Omega=(aI-\varepsilon bJ).\Omega=
(aF+\varepsilon b\mathcal{J}F)\otimes I
+\mathcal{J}(aF+\varepsilon b\mathcal{J}F)\otimes J
$$

\section{The action of $G$ and the nonlinear solutions}
We go back now to equations (1). We shall show that if $\Omega=F\otimes
I+\mathcal{J}.F\otimes J$ is a solution, then $\rho(\alpha).\Omega$ is also
a solution.  In order to do this in a coordinate free manner we shall
represent equations (1) in a different form independent on the Minkowski
metric. First we recall the general rule for multuplying vector valued
differential forms.  Let $\Phi=\alpha^i\otimes e_i, i=1,2,\dots,m$ and
$\Psi=\beta^j\otimes k_j, j=1,2,\dots,n$ be two $p$ and $q$ differential
forms on a manifold $M$ with values in the vector spaces $V^m$ and $V^n$ with
bases $\{e_i\}$ and $\{k_j\}$ respectively. Let $f_1:(\Lambda^p(M),
\Lambda^q(M))\rightarrow \Lambda^r(M))$ be a bilinear map, and
$f_2:(V^m,V^n)\rightarrow V^s$ be a bilinaer map. Now we can form the
expression $f_1(\alpha^i,\beta^j)\otimes f_2(e_i,k_j)\in\Lambda^r(M,V^s)$
which obviously is a $V^s$ valued $r$-form on M, and this last form we call a
$(f_1,f_2)$-product of $\Phi$ and $\Psi$.

Let now $M\equiv\mathbb{R}^4$. We recall now the Poincar\'e isomorphism
$\mathfrak{P}$ between the 2-forms and the 2-vectors built by making use of
the canonical volume forms
$\omega=\partial_x\wedge\partial_y\wedge\partial_z\wedge\partial_\xi$ and
$\omega^*=dx\wedge dy\wedge dz\wedge d\xi$:
\[
\mathfrak{P}(\partial_{x^i}\wedge \partial_{x^j})=
i(\partial_{x^j})\circ i(\partial_{x^i})\omega^*,
\]
where $i(X)$ is the substitution operator induced by the vector field $X$.
We make the composition $\mathfrak{D}=-\mathfrak{P}\circ \mathcal{J}$, which
is also an isomorphism. So, if $\alpha$ and $\beta$ are $p$ and $q$ forms
with $p\leq q$ we can form the $(q-p)$-form $i(\mathfrak{D}\alpha)\beta$.
Further, recalling our $\mathcal{G}$-valued 2-form $\Omega$, we choose the
symmetrized tensor product $\vee$ for a bilinear map:
$\vee:\mathcal{G}\times\mathcal{G}\rightarrow (\mathcal{G}\vee\mathcal{G})$.
We are ready now to compute $(\vee,i)(\Omega,\mathbf{d}\Omega)$; we obtain
\[
(\vee,i)(\Omega,\mathbf{d}\Omega)=\big[i(\mathfrak{D}F)\mathbf{d}F\big]
\otimes I\vee I+
\big[i(\mathfrak{D}(\mathcal{J}F))\mathbf{d}\mathcal{J}F\big]\otimes J\vee J+
\]
\[
\big[i(\mathfrak{D}F)\mathbf{d}\mathcal{J}F+
i(\mathfrak{D}(\mathcal{J}F))\mathbf{d}F\big]\otimes I\vee J.
\]
Clearly, the equation $(\vee,i)(\Omega,\mathbf{d}\Omega)=0$ gives equations
(1), where the Hodge $*$ is replaced with the complex structure
$\mathcal{J}$.

Let now $\alpha=(aI+\varepsilon bJ)\in G$. We have
$\rho(\alpha).\Omega=\rho(\alpha).F\otimes
I+\rho(\alpha).\mathcal{J}.F\otimes J$.  We shall show that if
$\Omega$ is a solution then $\rho(\alpha).\Omega$ is also a solution. In
fact,
\[
i(\mathfrak{D}(\rho(\alpha)F))\mathbf{d}(\rho(\alpha)F)=
a^2i(\mathfrak{D}F)\mathbf{d}F+
b^2i(\mathfrak{D}\mathcal{J}F)\mathbf{d}\mathcal{J}F+
\varepsilon ab\big[i(\mathfrak{D}F)\mathbf{d}\mathcal{J}F+
i(\mathfrak{D}\mathcal{J}F)\mathbf{d}F\big].
\]
It is seen that if $F$ satisfies (1) then the right side of the above
relation is equal to zero since it is a polynomial of the numbers $(a,b)$ with
coefficients equal to the three expressions staying in front of the
basis vectors $I\vee I, J\vee J$ and $I\vee J$ of
$(\vee,i)(\Omega,\mathbf{d}\Omega)$. Similar expressions are obtained for
$i(\mathfrak{D}(\rho(\alpha)
\mathcal{J}F))\mathbf{d}\rho(\alpha)\mathcal{J}F$ and for
$i(\mathfrak{D}\rho(\alpha)F)\mathbf{d}\rho(\alpha)\mathcal{J}F+
i(\mathfrak{D}(\mathcal\rho(\alpha){J}F))\mathbf{d}\rho(\alpha)F$.
Thus, $\rho(\alpha).\Omega$ is
also a solution. Moreover, since $a$ and $b$ are
constants, $(\rho(\alpha)\otimes \alpha).\Omega=(a^2+b^2).\Omega$ is also a
solution.  Hence, the above defined action of $G$ defines a symmetry in the
set of solutions to (1).

\section{Point dependent group parameters}
We are going now to see what happens if the group parameters become functions
of the coordinates: $a=a(x,y,z,\xi),\  b=(x,y,z,\xi)$. For convenience, we
shall write $\phi.\varphi\equiv u$ and $\phi.\sqrt{1-\varphi^2}\equiv p$.
Then (2) becomes
\begin{equation}
F=\varepsilon udx\wedge dz +
\varepsilon p dy\wedge dz+                       
u dx\wedge d\xi + p dy\wedge d\xi.
\end{equation}
Now, it can be readily checked [4] that $F$ will satisfy (1) iff in the
corresponding coordinate system $u$ and $p$ will satisfy the equation
\begin{equation}
L_\zeta\Big[det||\alpha(u,p)||\Big]=
(u^2+p^2)_\xi -\varepsilon(u^2+p^2)_z=0             
\end{equation}
where $L_\zeta$ is the Lie derivative with respect to the intrinsically
defined vector field $\zeta$.

Consider now the 2-form
\[ F_o=\varepsilon dx\wedge dz + dx\wedge d\xi,
\]
which, obviously, defines a (linear constant)
solution, and define a map $f:\mathbb{R}^4\rightarrow G$ as follows:
\[
\alpha(x,y,z,\xi)=f(x,y,z,\xi)=u(x,y,z,\xi)I+\varepsilon p(x,y,z,\xi)J,
\]
where the two functions $u$ and $p$ satisfy (7). Consider now the action of
$$
\rho(\alpha(x,y,z,\xi))
=u(x,y,z,\xi)\mathcal{I}+\varepsilon p(x,y,z,\xi)\mathcal{J}
$$
on
$F_o$. We obtain exactly $F$ as given by (6), i.e. the (linear) solution
$F_o$ is transformed to a nonlinear solution. This suggests to check if this
is true in general, i.e. if we have a solution (6) of (1) defined by the two
functions $u$ and $p$, and we consider a map
$\alpha:\mathbb{R}^4\rightarrow G$,
such that the components $a(x,y,z,\xi)$ and $b(x,y,z,\xi)$ of
$\alpha=a(x,y,z,\xi)I + \varepsilon b(x,y,z,\xi)J$ satisfy (7), then whether
the 2-form $\tilde F=\alpha(x,y,z,\xi).F= \big[a(x,y,z,\xi)\mathcal{I} +
\varepsilon b(x,y,z,\xi)\mathcal{J}\big].F$ will satisfy (1)?

For $\tilde F$ we obtain
\[
\tilde F=\rho(\alpha).F=(a\mathcal{I}+\varepsilon b\mathcal{J}).F=
\]
\[
\varepsilon (au-bp)dx\wedge dz +
\varepsilon (ap+bu)dy\wedge dz + (au-bp)dx\wedge d\xi
+(ap+bu)dy\wedge d\xi.
\]
Now, $\tilde F$ will satisfy (1) iff
\begin{equation}
\big[(au-bp)^2 + (ap+bu)^2\big]_\xi -
\varepsilon\big[(au-bp)^2 + (ap+bu)^2\big]_z=0. 
\end{equation}
This relation is equivalent to
\[
\big[(a^2+b^2)_\xi-\varepsilon(a^2+b^2)_z\big](u^2+p^2)+
\big[(u^2+p^2)_\xi-\varepsilon(u^2+p^2)_z\big](a^2+b^2)=0.
\]
This shows that if $F(u,p)$ is a solution to (1),
then $\tilde F(u,p;a,b)=\rho(\alpha(a,b)).F(u,p)$ will be a solution to (1)
iff the two functions $(a,b)$ satisfy (7), i.e. iff $\rho(\alpha(a,b)).F_o$
is a solution to (1). In other words: every nonlinear solution of (1),
$F(a,b)=\big[a\mathcal{I}+\varepsilon b\mathcal{J}\big]F_o$, defines a map
$\Phi(a,b):F(u,p)\rightarrow (\Phi F)(u,p;a,b)$, such that if $F(u,p)$ is
a solution to (1) then $(\Phi F)(u,p;a,b)$ is also a solution to (1).

\section{Interpretations}
The above considerations allow a structural interpretation of the set of
nonlinear solutions to (1). The whole set of nonlinear solutions to (1)
divides to subclasses $S(\vec{\bf r})$ of the kind (6) (every such subclass
is determined by the spatial direction $\vec{\bf r}$ along which the solution
propagates [4], it is the coordinate $z$ in our consideration). Every
 solution $F$ of a given subclass is obtained by means of the action of a
solution of equation (7) on the corresponding $F_o$:
$F(u,p)=\rho(\alpha(u,p)).F_o$. If $\alpha(u,p)\neq 0$ at some point of
$\mathbb{R}^4$ then we have $F^{-1}=\rho(\alpha^{-1}).F_o$ given by
$$
(F)^{-1}=\left(\frac{u}{u^2+p^2}\mathcal{I}
-\varepsilon\frac{p}{u^2+p^2}\mathcal{J}\right).F_o .
$$
In view of these remarks equation (5) acquires a natural interpretation of
{\it eigen} equation for the operator $\rho(\alpha)\otimes\alpha$: every
$\Omega$ in a given subclass is an eigen state for any
$\rho(\alpha)\otimes\alpha$ since all states $\Omega$ of this subclass
describe propagation along the same spatial direction, or along the
4-dimensional direction defined by the only isotropic eigen vector
$\zeta$ of the energy-momentum tensor, and the eigen value $(a^2+b^2)$ is
just the energy density in the point dependent case.

Further, the relation
\begin{equation}
\rho(\alpha).\Omega_o
=\rho(\alpha).(F_o\otimes I+\mathcal{J}.F_o\otimes J)
=F\otimes I+\mathcal{J}.F\otimes J=\Omega
\end{equation}
suggests to consider $\Omega_o$, as a "vacuum state", and the action of
$\rho(\alpha)$ on $\Omega_o$ as a "creation operator", provided
$\alpha(x,y,z,\xi)\neq 0, (x,y,z,\xi)\in D\subset\mathbb{R}^4$, satisfies (7)
in $D$.  Since $\rho(\alpha).\rho(\alpha^{-1}).\Omega_o= \Omega_o$ we see
that every such "creation operator" in $D$ may be interpreted also as an
"annihilation operator" in $D$.

Clearly, every solution $\Omega=\rho(\alpha).\Omega_o$ may be represented in
various ways in terms of other solutions to (7) in the same domain $D$., i.e.
we have an example of a {\it nonlinear} "superposition".  Also, making use of
the {\it trigonometric} representation of $\alpha$, we see that every
solution in a natural way acquires the characteristics {\it amplitude}
$\phi=|\alpha|=\sqrt{a^2+b^2}$ and {\it phase} $\psi=arccos(\varphi)$ (see
relation (2)).  Moreover, every $F=\rho(\alpha).F_o$ generates
$(F)^n=\rho(\alpha ^n).F_o$ and $\sqrt[n]{F}=\rho(\sqrt[n]{\alpha}).F_o$
through the well known Moivre formulae, and in this way we obtain new
solutions of the same subclass.

If we consider the trivial bundle $\mathbb{R}^4\times G\rightarrow
\mathbb{R}^4$ then every such subclass of nonlinear solutions is generated by
a subclass of sections $\alpha(u,p)$ where $(u,p)$ satisfy equation (7).
Explicitly, the solution is obtained in the form of $\alpha(x,y,z,\xi).F_o$.
Hence, instead of a linear structure, every such subclass of nonlinear
solutions appears rather as an "orbit" of an action upon $F_o$
of those sections satisfying equation (7).

Another look at the situation is to identify $\mathcal{G}$ with the field of
complex numbers $\mathbb{C}$ and to consider the trivial bundle
$\mathbb{R}^4\times \mathbb{C}$. Then, considering those nonsingular sections
$Z=(a,b)$ of this bundle the squared modules $|Z|^2=(a^2+b^2)$ of which
satisfy (7), we see that the point-wise multiplication of these sections
sends two solution-sections $Z_1=(a,b)$ and $Z_2=(u,p)$ to the
solution-section $Z_1.Z_2=(au-bp,ap+bu)$.

\section{Conclusion}
This paper clearly shows that, the EED vacuum equations, as well
as Maxwell vacuum equations, do {\it not necessarily} use
pseudoeuclidean metric structures, the basic object used in both systems of
equations is the complex structure $\mathcal{J}$.  The whole set of nonlinear
solutions to equations (1) divides to subclasses, and each such subclass is
determined by the direction of propagation in the 3-space, or by the
corresponding eigen vector $\zeta$. The elements of each subclass are
obtained by means of the action on a specially chosen 2-form $F_o$ of those
(point-dependent) elements of $G$ which satisfy equation (7), i.e. which have
invariant with respect to $\zeta$ determinant. So, every two nonlinear
solutions determine another nonlinear solution provided their domains have
non-empty section, i.e.  we have a map inside a given subclass.  Every
nonlinear solution $F(u,p)$ acts as a symmetry transformation in its
subclass, and we have $F_1(a,b).F_2(u,p)=F_2(u,p).F_1(a,b)$. Clearly, this
structure of the subclass may be considered as generated by the usual
multiplication of the complex numbers.

If we consider spatially finite elements $\tilde\alpha(x,y,z,\xi)$,
satisfying (7), we obtain $(3+1)$ soliton-like "creation operators"
$\rho(\tilde\alpha)$, which produce $(3+1)$ soliton-like solutions
$\rho(\tilde\alpha).\Omega_o$ through acting on the "vacuum" state
$\Omega_o$. So, the two solitons $\rho(\tilde\alpha)F_o$ and
$\rho(\tilde\alpha^{-1})F_o$ kill one another. All of these solutions have,
of course, well defined {\it amplitudes} and {\it phases}. If we have a
subclass of such soliton-like solutions with mutually non-overlapping 3d
domains we obtain a flow of photon-like objects.

\vskip 1cm {\bf REFERENCES}
\vskip 0.5cm

[1] S. Donev, M. Tashkova, Proc. Roy. Soc. of London A 450, 281 (1995).

[2] J. Synge, {\it Relativity: The Special Theory}, North-Holland, 1958.

[3] S. Donev, LANL e-print:  math-ph/0106008.

[4] S. Donev, M. Tashkova, Proc. Roy. Soc. of London A 443, 301 (1993).

\end{document}